\documentclass[pre,superscriptaddress,showpacs,amsmath,amssymb]{revtex4}

\usepackage{epsfig}

\usepackage{bm}
\usepackage{graphicx}
\usepackage{amsmath}
\usepackage{amssymb}
\graphicspath{{Figures/}}
\usepackage[french]{babel}
\begin{document}

\title{Aging mechanism in tunable Pickering emulsion}
\author{S. Fouilloux, A. Thill, J. Daillant(*) and F. Malloggi}
\affiliation{Laboratoire Interdisciplinaire sur l'Organisation Nanom\'etrique et Supramol\'eculaire (LIONS),\\
3685 CEA NIMBE-LIONS,\\
CEA Saclay, F-91191 Gif-sur Yvette, France.\\
(*) Presently at Soleil Saclay.
}

\date{\today}

\begin{abstract}
We study the stability of a model Pickering emulsion system. A special counter-flow microfluidics set-up was used to prepare monodisperse Pickering emulsions, with oil droplets in water. The wettability of the monodisperse silica nanoparticles (NPs) could be tuned by surface grafting and the surface coverage of the droplets was controlled using the microfluidics setup. A surface coverage as low as 23$\%$ is enough to stabilize the emulsions and we evidence a new regime of Pickering emulsion stability where the surface coverage of emulsion droplets of constant size increases in time, in coexistence with a large amount of dispersed phase. Our results demonstrate that the previously observed limited coalescence regime where surface coverage tends to control the average size of the final droplets must be put in a broader perspective.
\end{abstract}

\pacs{47.10.+g, 47.57.Bc, 47.57.J-}

\maketitle

%_______________________________________________________________________________
\section{Introduction}
Ramsden \cite{Ramsden1903} and Pickering \cite{Pickering1907} observed one century ago that solid particles are able to stabilize emulsions, now referred to as Pickering emulsions \cite{Binks2000}. Depending on the particles wettability, either O/W emulsions (hydrophilic particles, contact angle $\theta < 90$\degre) or W/O emulsions (hydrophobic particles, $\theta > 90$\degre) are preferably stabilized. The attachment energy of the nanoparticles at the interface is normally large and gets stronger when the nanoparticles get larger. It is for example usually admitted that NPs of more than 10 nm are irreversibly adsorbed at the oil water interface \cite{Binks2001}. 
Beyond their stability, another attractive characteristic of Pickering emulsions is their monodispersity which results from an instability called "limited coalescence". The total amount of particles initially adsorbed is usually not sufficient to efficiently protect the oil-water interfaces. In such a case, the emulsion droplets coalesce leading to a progressive reduction of the total interfacial area between oil and water interfaces. As particles are irreversibly adsorbed, the reduction in oil/water interfacial area goes along with an increase in surface coverage. The coalescence continues until a surface coverage sufficient to protect the oil/water interface is reached \cite{Arditty2003}.

\begin{figure} [h]
\centering
\includegraphics[scale=0.4]{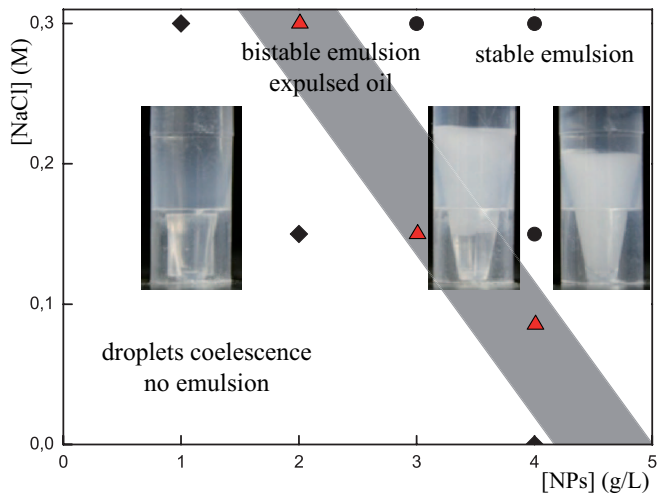}
\caption{Phase diagram as a function of the ionic strength and NPs concentration.}
\label{PhaseDiagram}
\end{figure}

To progress in our understanding of the stability behavior of Pickering emulsions, it is therefore important to prepare droplets of both controlled size and controlled surface coverage using well defined NPs as stabilizers. The past decade has seen the rise of microfluidic tools, see for a non-exhaustive review \cite{Squires2005}, that allow, among other, a good control of emulsions, i.e. monodisperse droplets \cite{Garstecki2005} and surface coverage by particles \cite{Studart2009} \cite{Kim2008}. In this article we have set-up a microfluidic system allowing such a coupled control. We produce monodisperse oil-in-water (O/W) droplets stabilized by monodisperse silica NPs. The surface coverage of the droplets by the NPs is controlled through the microfluidic process. Due to the fine controlled of NPs coverage, we are able to determine the stability diagram of the Pickering emulsion (see Fig.\ref{PhaseDiagram}). In addition, we evidence a new stability regime where a large amount of dispersed phase is expulsed while the other oil droplets keep their native size. 
%
%_______________________________________________________________________________
\section{Experimental set-up}
\subsection{Functionalized nanoparticles}

Well-controlled silica nanoparticles were prepared according to the method of Fouilloux \textit{et al.} \cite{Fouilloux2010} and their surface energy was modified by grafting trimethylethoxysilane (TMES) (Supplementary Material: Nanoparticles preparation). Their size and aggregation state was then assessed with SAXS (Small Angle X-ray Scattering) and DLS (Dynamic Light Scattering) after dialysis. Bare silica NPs have 4.5 silanol/nm$^2$ and are reduced up to 1 silanol/nm$^2$ after grafting TMES. Surface density of silanol functions is determined by titration (total number of silanol functions) and SAXS measurements (total surface of NPs). Using DLS size measurements, it is observed that up to a total TMES concentration of 4.6 TMES molecules/nm$^2$ ($\sim$ 3 silanol/nm$^2$), the surface modified NPs are still well dispersed in water (Supplementary Material: Nanoparticles titration and Surface modification of NPs).  SAXS measurements indicate that a repulsive interaction exists between the NPs even after the surface treatment at 4.6 TMES/nm$^2$. When more than about 8 TMES/nm$^2$ is used in the surface modification reaction, the NPs are no longer stable in water, but it is still possible to partially disperse the NPs using ethanol as a solvent. In the following we only used NPs dispersed in water.
\begin{figure*} [b]
    \begin{center}
        \includegraphics[width=0.85 \textwidth]{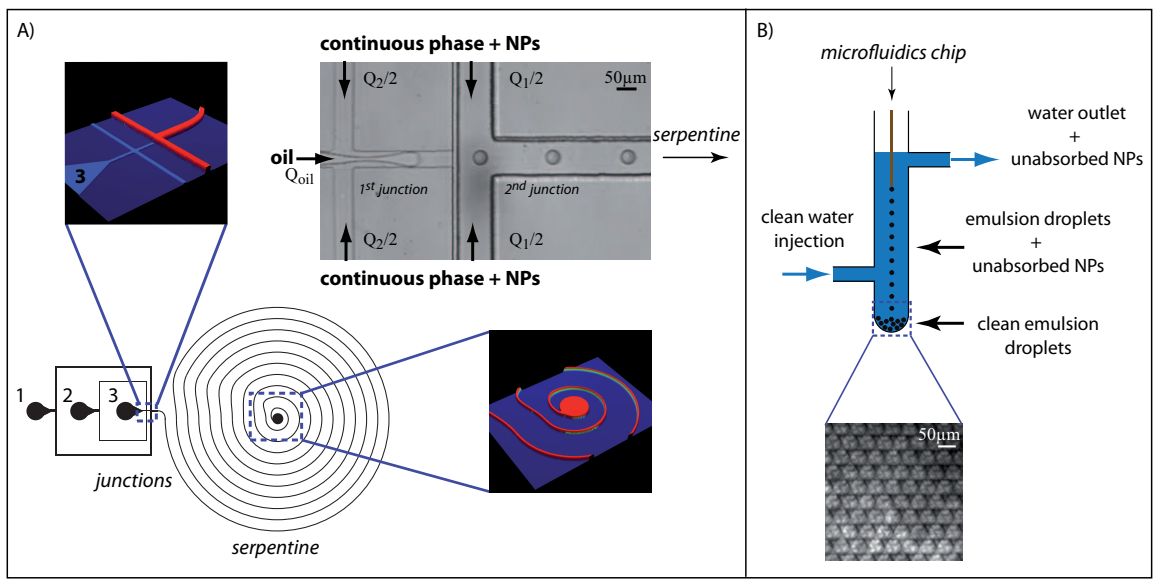}
    \end{center}
\caption{\textbf{A}. Microfluidics set-up - Emulsion generator. Flow rates used for the emulsion conditions: Q$_{oil}$ = 0.3$\mu$l/min – Q$_1$= 7$\mu$l/min – Q$_2$= 1.5$\mu$l/min. \textbf{B}. Millifluidics set-up - Emulsion cleaner.}
\label{Fig01} 
\end{figure*}

\subsection{Microfluidic droplets generator}
The microfluidic device consists of two flows focusing in series (see Fig.\ref{Fig01}.A) microfabricated using standard multilayers soft technology \cite{Xia1998}\cite{Ziaie2004}. The first junction (width 50$\mu$m and height 11$\mu$m) generates the oil droplets: the dispersed phase (fluorinated oil) is sheared by the continuous phase (water + NPs) until the interface’s destabilization. Once the droplets are formed they flow to the second junction. In order to optimize the NPs adhesion the second junction (two steps lithography \cite{Malloggi2010}) has a larger width (100$\mu$m) and height (95$\mu$m): droplets adopt a 3D shape (sphere of radius 22$\mu$m) and are fully surrounded by NPs. To maximize the contact time of NPs with the oil droplets, we make a 40 cm long serpentine which, according to the flow rate we fixed ($Q_{oil}$=0.3$\mu$l/min, $Q_{cont.phase}$ = 8.5$\mu$l/min), gives a resident time of 27s. The Stokes-Einstein formula gives a diffusion coefficient D = 3.1 10$^{-11}$ m$^2$.s$^{-1}$ (NPs of radius 7.1 nm) and the mean transversal diffusion length writes $l = \sqrt{2Dt}=40 \mu m$. Hence almost all the NPs entered at the 2nd junction reach the droplet’s interface at the end of the serpentine (the maximal lateral diffusion length is 50$\mu$m when no droplet is in the channel and 30$\mu$m with droplets). Another parameter to adjust is the number of NPs in contact with the oil droplets. We used concentrations of NPs that range from 0.4 to 3.5 g.L$^{-1}$. In many cases, a NaCl salt concentration of up to 0.3M was used to screen the repulsive electrostatic repulsions between the NPs. We checked that NPs do not aggregate at such a high ionic strength (data not shown). At the outlet of the microfluidic chip, only part of the injected NPs are adsorbed at the oil interface. The emulsion droplets coexist with a suspension of unabsorbed NPs. In order to remove the excess NPs, the outlet of the microfluidics is connected to a rinsing millifluidic set-up (Figure \ref{Fig01}-B). It consists of a homemade glass vial with one inlet (linked to the microfluidics device) and two lateral connections. A pure Millipore water flows from bottom to top through the lateral connections. The emulsion droplets enter from the top of the vial and fall in a clean water counter flow. The fluorinated oil density -1.82 $g/cm^3$- guarantees that the droplets will settle even in a significant counter flow. \textbf{The non-adsorbed NPs are thus removed from the emulsion suspension}. At the bottom of the set-up, only the NPs previously adsorbed at the interface of the oil droplet will remain (see the emulsion crystal-like structure in Figure01-B). Hence the use of variable NPs concentrations allows producing Pickering emulsions with a tunable surface NPs coverage.

\subsection{Pickering emulsion}
To produce the emulsions in the microfluidic device, we have selected a surface modification which guarantees a colloidal stability in water. The NPs with 70$\%$ residual silanol surface groups ($\sim$3 silanol/nm$^2$) were introduced at concentrations of 1, 1.7 and 3.5 g.L$^{-1}$ in the presence of 0.3M NaCl. This salt concentration is enough to almost completely screen the repulsive interaction between the NPs. However, the aggregation kinetic of the NPs is very slow compared to the residence time in the fluidic system ($\sim$30s). This guarantees that only isolated NPs are in contact with the oil/water interface in the chip. According to the droplet emulsion conditions ($Q_{oil}$=0.3$\mu$ l/min and radius=22$\mu m$) the oil surface produced is 4 10$^{-5}$ m$^2$/min. At the lowest NPs concentration of 0.4 g.L$^{-1}$, the available NPs cross section is 1.6 10$^{-4}$ m$^2$/min. Therefore the NPs available to cover the oil droplets are at least 4 times larger than the oil surface created within the emulsion. The diffusion-convection characteristics of the NPs in the microchannel flow together with the residual electrostatic repulsive interactions however prevent from reaching a full coverage of the droplets at the chip outlet for the lowest concentration. The NPs concentration is thus introduced in large excess in order to force significant adsorption within the available contact time and the excess NPs is removed in the millifluidic counter flow device.
\subsection{Surface coverage measurements}
\begin{figure} 
\centering
\includegraphics[scale=0.75]{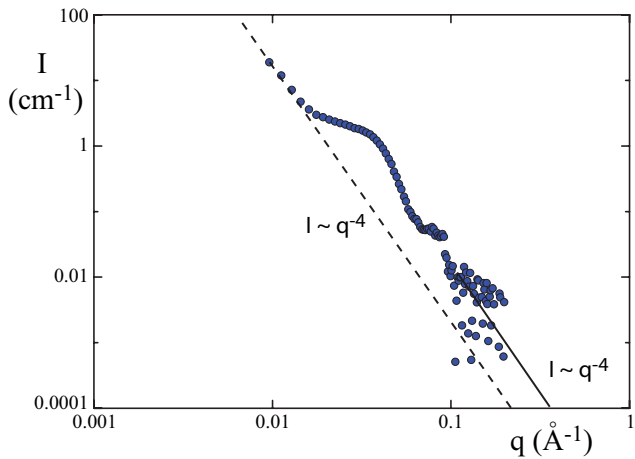}
\caption{SAXS measurements of an oil-in-water emulsion stabilized with monodisperse silica NPs. The excess NPs have been removed by counter flow rinsing before collection of the droplets. The full line represents the Porod extrapolation of the NPs and the dotted line the Porod extrapolation of the oil droplets.}
\label{Fig02}
\end{figure}

Some emulsions were collected directly in a glass capillary for SAXS analysis which is a powerful technique to probe the size, shape, polydispersity and concentration on NPs \textit{in situ} \cite{Boukari1997}. The capillary is connected to the bottom of the millifluidic rinsing device: the emulsion collected in the SAXS capillary only contain partially covered oil droplets without an excess of free NPs. Figure \ref{Fig02} shows SAXS measurements for an emulsion prepared with the 7.5 nm NPs having 70$\%$ residual silanol groups on their surface at a concentration of 1 g.L$^{-1}$ in the presence of 0.3M NaCl. 
In the very low q range (q$\sim$0.01${\AA}^{-1}$), the scattering intensity is explained by the oil/water interface. The amplitude of this intensity is proportional to the surface of the emulsion droplets, to the scattering length density contrast between the two components of the interface and to the NPs surface coverage. The amplitude writes:
\begin{equation}
I(q)q^4=2\pi\Delta\rho^2\frac{S}{V}
\end{equation}
where $\Delta\rho$ is the scattering length density contrast between oil and water, $S$ is the surface of the oil/water interface and $V$ is the scattering volume. From this equation, it is possible to obtain the total surface of the oil droplets.

In the intermediate q range, the scattered intensity is due to the presence of the adsorbed NPs. At the large q range (q$>$0.1${\AA}^{-1}$), despite the scattered signal, there is a characteristic Porod law which is proportional to the quantity of NPs present at the interface.
SAXS analysis allows an accurate coverage determination of oil droplets by NPs (for data treatment details see Supplementary Material: Analysis of SAXS measurements).  
%_______________________________________________________________________________
\section{Results}
\begin{figure} [b]
\centering
\includegraphics[scale=0.75]{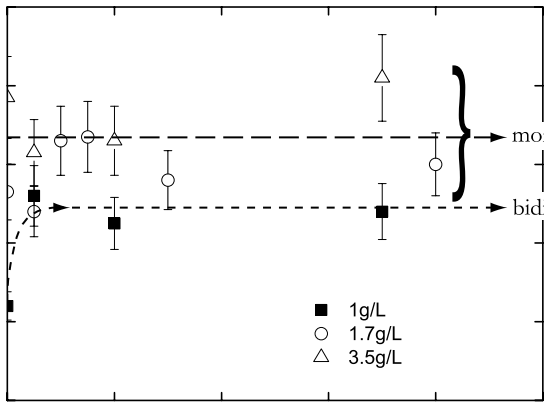}
\caption{Surface coverage of the microfluidic prepared emulsions as a function of aging time for three NPs initial concentrations.}
\label{Fig03}
\end{figure}

\begin{figure} 
\centering
\includegraphics[scale=1]{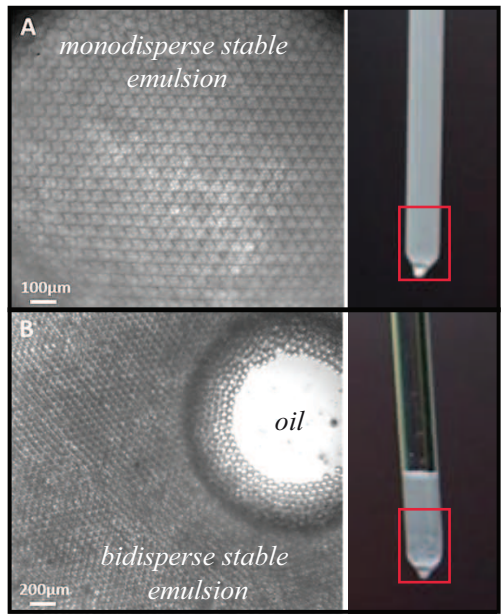}
\caption{Microscope image of the Pickering emulsions after one day aging. \textbf{A}. Initial surface coverage $>$50$\%$. \textbf{B}. Initial surface coverage $\sim$25$\%$.}
\label{Fig04}
\end{figure}

The Figure \ref{Fig03} shows the obtained surface coverage for three different initial NPs concentrations (1, 1.7 and 3.5 g.L$^{-1}$) in the presence of 0.3M NaCl, which were measured immediately after preparation and also after several days of aging time up to about one week. The initial surface coverage of the oil droplets appears to be significantly dependent on the NPs initial concentration. The surface coverages measured immediately after the sample preparation are 24 $\pm$ 3 $\%$ for 1g.L$^{-1}$, 53 $\pm$ 6.6 $\%$ for 1.7 g.L$^{-1}$ and 77 $\pm$ 9.6 $\%$ for 3.5 g.L$^{-1}$. The contact time with the NPs is the same for all concentrations: the surface coverage is efficiently controlled by the NPs concentration in the continuous phase. The need to use a large excess of NPs to control the coverage is not a problem when the excess NPs is removed before analysis. Experiments (not shown) were the excess NPs is not removed immediately show an equivalent coverage of $\sim$80$\%$ and stable emulsions whatever the NPs concentrations.
The rinsed emulsions having various initial surface coverage show interesting aging behavior. The emulsions prepared with the highest NPs concentration have an initial surface coverage of $\sim$75$\%$. This coverage appears to be stable over time. The emulsion appears also stable macroscopically.
On the contrary, a change in the surface coverage is observed for emulsions prepared with the lowest concentrations experiment.  The initial coverage is slightly above 20$\%$ but it rapidly increases up to about 50$\%$ after only 1 day. Macroscopically, the emulsion is not as stable as the one prepared with a high NPs concentration: we observe an excess of oil (big droplets) coexisting with the native emulsion droplets. The Figure \ref{Fig04} shows such a rinsed emulsion observed with an optical microscope one day after the formation. It is clear that for emulsions with initial surface coverage above 50$\%$ (Fig.\ref{Fig04}-A) oil droplets are stable whereas in the case of initial surface coverage of 24$\%$ (Fig.\ref{Fig04}-B) big oil droplets co-exist with native one. This excess oil was not present after the emulsion rinsing step. 

%
%_______________________________________________________________________________
\section{Discussion}
 In the following we discuss this new result and we propose a scenario for this excess oil expulsion. At the outlet of the microfluidic chip, all droplets have the same diameter. If we neglect Oswald ripening and if we assume that NPs are conserved upon coalescence (i.e. no NPs are expulsed from the interface), we write the droplet size emulsion $R_n = R_i \sqrt[3]{n}$, the number of adsorbed NPs $N_n=n4\pi{R_i}^{2}\tau_i$ and the surface coverage $\tau_n=\tau_i \sqrt[3]{n}$ as a function of the initial droplet size $R_i$, the initial surface coverage $\tau_i$ and the number of coalescence events $n$. 
According to the results shown is Fig.\ref{Fig03}, when 1 g.$L^{-1}$ NPs suspension is used to stabilize the droplets, the initial surface coverage $\tau_i$ is 23$\%$. After one day of aging, the surface coverage increased up to 50$\%$ and stays almost constant. If we assume no NPs expulsion from the oil/water interface, the final droplet size would writes $R_n=R_i\tau_n/\tau_i$  which gives $R_n$ = 43 $\mu m$. This scenario is not what we have as confirmed by the pictures on Fig.\ref{Fig04}-B. Indeed,we see particles with the initial size ($\sim$20$\mu m$). Hence some NPs have been added to the surface of droplets without coalescence events. As excess oil has appeared, we make the following hypothesis to explain our observation. The 23$\%$ surface coverage could be very close to a perfect stabilization of the droplets, but for such coverage it could remain rare coalescence events. Contrarily to the limited coalescence, we suppose that a significant part of the NPs adsorbed on the droplets are released in the suspension upon coalescence. In such a scenario, coalescence may not be a stabilizing mechanism but could become a catastrophic event which further increases the coalescence probability for the droplet if the surface coverage after coalescence is less than before. 

To go a step further in the destabilization mechanism comprehension, we make an energetic balance between adhesion energy and gain in interfacial energy to see if our assumption is feasible.

\textbf{NPs Energy adhesion.} The adsorption of partially wetted NPs is often believed to be irreversible as soon as their size is close to 10 nm or more. The adhesion energy per NPs is indeed roughly given by $E_{ad} =\pi r^2 \gamma (1 \pm cos\theta)^2$.  In this study, we have used silica NPs having a size of 7.1 nm and at least 70$\%$ of their silanol surface sites. Binks \textit{et al.} \cite{Binks2002} have proposed computed contact angle as a function of the residual silanol groups at the surface of silica NPs. For 75$\%$ SiOH, they predicts a contact angle of 65\degre for $\gamma$ =50 mN/m. For a NP of 7.1 nm, $\gamma$=50 mN/m and $\theta$=65\degre, this gives almost $E_{ad} =700 kT$. According to this estimation it is clear that, once adsorbed at the oil water interface, NPs will not escape through only thermal activation.

\textbf{Energy released during coalescence events.} Upon coalescence the droplet released part of their interfacial energy due to the decrease of oil/water interface. For two droplets having a radius R, the variation of interfacial energy, if a constant surface tension is assumed, writes:
\begin{equation}
\Delta E=(2^{2/3}-2)4\pi R^2 \gamma
\end{equation}
If we compare this excess interfacial energy per adsorbed NPs to the adhesion energy, we obtain the ratio:
\begin{equation}
\alpha =\frac{\Delta E}{N_{ad}E_{ad}}=\frac{2^{-1/3}-1}{(1-cos\theta)^2\tau}
\end{equation}
For NPs having a contact angle of 65\degre, the energy ratio is simply $\alpha\sim -0.62/\tau$ With an initial surface coverage of $\tau$= 0.23, the gain of interfacial energy per NPs is 2.7 times more than the adhesion energy. The surface coverage stabilizes at $\tau\sim$0.65-0.7 which corresponds to the point where the interfacial energy gain per NPs upon coalescence is no longer superior to the adhesion energy. This simple energetic comparison is just intended to show that the assumption of a NPs release upon coalescence of a Pickering emulsion may not be an impossible process. Surprisingly in this case, this simple calculation seems also to correspond to the stabilization process of the droplets. However the precise energetic transferred to the NPs remain unclear as part of the energy is used to move fluids. In our case, the fluorinated oil has a very low viscosity which is favorable to transfer more energy to the NPs. 
%_______________________________________________________________________________
\section{Conclusion}
In conclusion, we have built a new and original coupled microfluidic/millifluidic system to prepare model Pickering emulsions. The droplets are stabilized by monodisperse NPs having controlled surface properties. This set-up allows tuning the surface coverage of the final emulsion by changing the NPs concentration in the microfluidic system. The obtained surface coverage is experimentally determined in situ using SAXS measurements of the as prepared emulsion. This technique does not require any manipulation of the sample. It has been thus possible to follow the aging of emulsions having varying initial surface coverage from 23$\%$ to 77$\%$.  We have identified two stability scenarios depending on the initial emulsion surface coverage. For the lowest surface coverage ($<$50$\%$) a new and original destabilization regime is observed. An excess disperse phase coexists with almost undisturbed droplets. We assume that this could be due to a coalescence induced release of NPs. We argue that this process is energetically plausible and that the comparison of the interfacial energy release par adsorbed NPs compared to their adhesion energy could be a guide to understand Pickering emulsion stabilization and aging.

ACKNOWLEDGMENTS.
This project has benefited from the support of the Cnano Ile de France grant SINSEM. The Ile de France region is acknowledged for its support of the SEM apparatus used for the NPs imaging and Mathieu Pinault and Aurélie Habert for their help. 

\section{Supplemantary Material: Aging mechanism in tunable Pickering emulsion}
%_______________________________________________________________________________
\subsection{Nanoparticles preparation}
The chemicals used for the NPs synthesis are TetraEthylOrthoSilicate (TEOS), l-Arginine, chloride acid 0.1M and ethanol. They were supplied by Aldrich and used as received. High-purity deionized water (18.2 MΩ.cm) was produced using Millipore Milli-Q Gradient system. The synthesis follows the same protocol as in our previous work \cite{Fouilloux2010}. It is performed in a 2000 ml reactor thermostated with hot water circulation at 60\degre C and agitated by a blade mixer. The vapors are cooled in a water condenser. 1000 ml of a 6 mM l-Arginine solution is first introduced in the reactor. Once the solution reaches the desired temperature, 80 mL of TEOS is added in the reactor and form an organic phase on top of the aqueous solution. The stirring rate is fixed such that the two phases are well mixed and form an emulsion during the synthesis (1500 rpm). The reaction is kept at constant stirring speed and temperature during at least 24h and then allowed to cool down to ambient temperature.
The surface modification of the particles is performed after determination of their size and concentration (i.e.their total exposed surface). A 200 mL volume of suspension is placed in a vial and agitated by a magnetic stirrer (750 rpm). A volume of 0.5 to 4 mL of trimethylethoxysilane (TMES) is added, corresponding to 2.3 to 18.2 TMES molecules/nm$^2$ of particle surface and the grafting reaction is performed during 4 to 24 h.
Once the reaction is complete, the particles are washed 3 to 5 times in a Millipore Amicon stirred cell (30 kDa membrane) to remove the arginine and reaction by-products. The particles are washed and redispersed in water or ethanol depending on their hydrophobicity: particles grafted with more than 9 TMES molecules/nm$^2$ aggregate in water but redisperse readily when washed in absolute ethanol.
The size and aggregation state of the NPs is assessed with SAXS and DLS after dialysis.

%_______________________________________________________________________________
\subsection{Nanoparticles titration}
The particles hydrophobicity is characterized via the titration of the residual silanol functions on their surface. Approximately 0.2 g of particles are dried by solvent evaporation in an oven at 90\degre C during 24h and then dispersed in 16 mL of a 20 g.L$^{-1}$ NaCl solution. The titration is performed with a Hanna automatic titrator, after degassing the solution by nitrogen bubbling during 30 min. After each addition of the 0.1M HCl titrating solution, the sample is allowed to equilibrate for at least 2 min before the pH is measured. The titration is stopped when the pH reaches 2. The volume necessary to lower the pH of the solution from 6.5 to 2.5 is recorded. A blank experiment is performed on a nanoparticle-free NaCl solution and subtracted from this value. Since the total surface of the particles is known from SAXS measurements, this titration enables the determination of the total number of silanol functions in the sample; hence the surface density of silanol functions.

%_______________________________________________________________________________
\subsection{Surface modification of NPs}
\begin{figure} [h]
\centering
\includegraphics[scale=0.4]{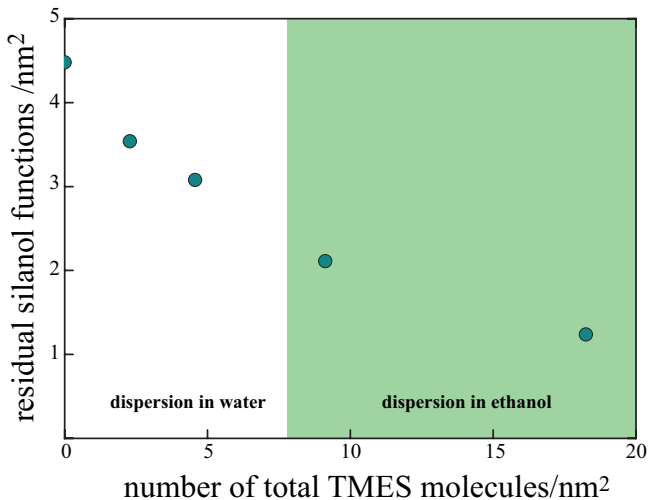}
\caption{ Titration of the residual silanol on NPs surface as a function of contacted silane molecules. }
\label{FigS1}
\end{figure}

The surface modification of the NPs is assessed through titration after dialysis. Figure \ref{FigS1} shows the quantity of silanol function per surface unit of NPs as a function of the total volume of TMES introduced during the surface modification reaction. It is observed that the more TMES introduced, the less residual silanol groups are present on the surface. Bare silica NPs have 4.5 silanol/nm$^2$. This value is reduced to about 2 silanol/nm$^2$ when 8 TMES molecules per nm$^2$ are introduced in the reaction media. Although the surface modification reaction has a reduced yield between 25-30$\%$, it appears possible to tune the residual silanol groups density efficiently down to about 1 silanol/nm$^2$ when 18 TMES molecules per nm$^2$ are used. We clearly noticed a change of NPs hydrophilicity since with more than 8 TMES molecules/nm$^2$ the NPs do no longer disperse properly in water but in ethanol. These modified surface NPs are then measured by SAXS and DLS.
%

%_______________________________________________________________________________
\subsection{NPs aggregation}
The figure S2 shows the DLS size as a function of TMES molecules/nm$^2$ and SAXS curves corresponding to the bare and surface modified NPs with $\sim$3 silanol/nm$^2$ (4.56 TMES molecules/nm$^2$). On the DLS size measurements, it is observed that up to a total TMES concentration of 4.6 TMES molecules/nm$^2$, the surface modified NPs are still well dispersed in water. The SAXS curves give two useful information. First, the intensity oscillations at large angle are still present after the surface treatment. There is no noticeable shift of these oscillations which shows that the size of the NPs is the same.  We also observe the presence of an interaction peak at small angle for both samples. This indicates that a repulsive interaction exists between the NPs even after the surface treatment at 4.6 TMES/nm$^2$. It is thus possible to maintain the NPs stability while modifying their hydrophobicity. When more than about 8 TMES/nm$^2$ is used in the surface modification reaction, DLS measurements become impossible as the NPs are no longer stable in water. In these cases, it is still possible to partially disperse the NPs using ethanol as a solvent. DLS measurements of the NPs redispersed in ethanol show that the NPs are not fully aggregated. The measured size is slightly different from the one obtained in water indicating that the NPs are probably in the form of small aggregates (50nm instead of 20nm for dispersed NPs). 

\begin{figure} [b]
\centering
\includegraphics[scale=0.75]{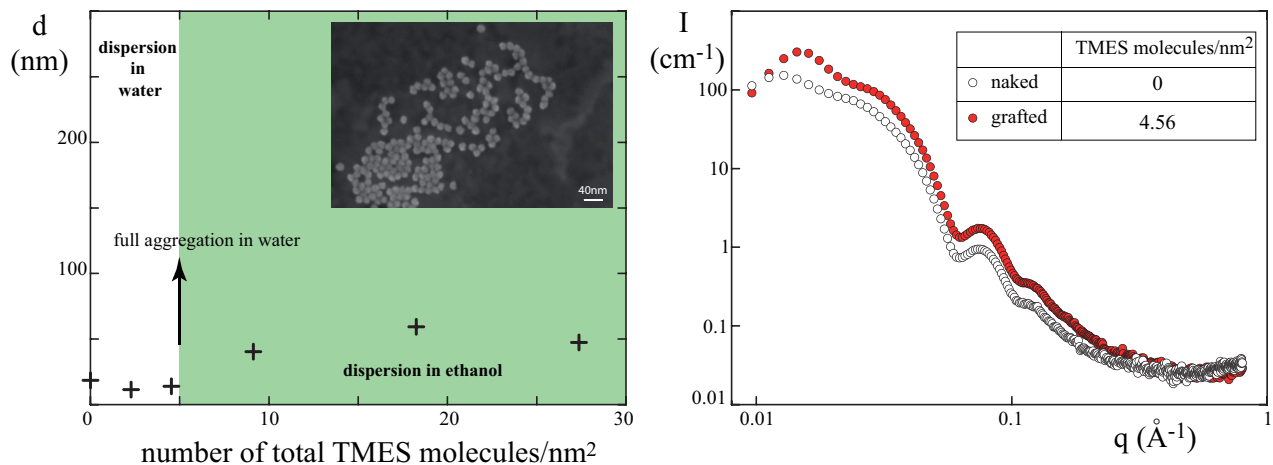}
\caption{ Characterization of the NPs used for the emulsion preparation. Left. DLS measurements of functionalized NPs. Inset TEM (or SEM) picture of bare silica nanoparticles. Right. SAXS measurements of bare and grafted NPs.}
\label{FigS2}
\end{figure}
%

%_______________________________________________________________________________
\subsection{Microfluidics droplet generator and Millifluidics emulsion cleaner}
Microchannels were microfabricated using standard multilayer soft technology \cite{Xia1998}\cite{Ziaie2004}\cite{Malloggi2010}. The polydimethylsiloxane (PDMS) elastomer (Sylgard 184, Dow Corning) was poured and further cured on a SU-8 mold (Microchem). Holes were punched for the inlets and the PDMS replicates were sealed to glass slides following oxygen plasma treatment.
%_______________________________________________________________________________
\subsection{Small Angle X-ray Scattering}
Small Angle X-ray Scattering (SAXS) is a powerful technique to probe the size, shape, polydispersity and concentration of nanoparticles in situ \cite{Boukari1997}. The scattered intensity I(q) is measured as a function of the scattering wave vector $q=(4πsin(\theta/2))/\lambda$ where $\lambda$ is the wavelength of the incident X-ray beam (λ=1.54${\AA}$). The measurements are performed on a laboratory SAXS apparatus described by Zemb et al \cite{Boukari1997}. Since this publication, the following modifications have been performed. The X-ray source is a copper rotating anode operated at 3 kW using a micro focused source. We use a multilayer Xenocs mirror to obtain a monochromatic parallel X-ray beam. The beam is collimated using three slits under vacuum. The sample is placed at 230 cm after the mirror and at 122 cm from the detector plane. A beam stopper is placed under vacuum before the detector. The detector is a Mar300, an automatic imageplate based system from Marresearch.
The sample measurements are performed using a specially designed sample cell that allows variable beam path length. The sample thickness is first set to 1 mm and a scattering picture is obtained on the detector after an accumulation time of 3600s. Then a second X-ray scattering image is obtained for the empty sample cell on the detector for the same counting time. The two pictures are radially averaged and normalized with standard procedures to give the scattered intensity of the sample (in cm$^{-1}$).

%_______________________________________________________________________________
\subsection{Analysis of SAXS measurements}
Due to the different size scales involved, with the present SAXS measurement, it is possible to obtain the NPs droplets surface coverage using general scattering properties. First the X-ray beam is attenuated through the sample. This beam attenuation depends on the sample thickness and composition. We have:
\begin{equation}
T= e^{-\sum f_i \mu_i e}
\end{equation}
where $T$ is the sample transmission, $f_i$ and $µ_i$ are respectively the mass fraction and the linear absorption coefficient of the sample component (water, oil, NPs). It is reasonable to assume that the mass fraction of NPs after the droplets rinsing is negligible compared to oil and water as the initial mass fraction is less than 0.3$\%$. Even if a complete coverage of the settled droplets is envisaged, the maximum possible mass fraction will not be in excess of 0.5$\%$ due to the large size difference between the droplets (40 $\mu$m) and the NPs (7.1 nm). Assuming that the NPs mass fraction is negligible compared to the water and oil fractions, it is possible to obtain the oil mass fraction $f_o$ from the transmission measurements:

\begin{equation}
f_o=\frac{-ln(T)-e\mu_w}{(\mu_o-\mu_w)e}
\end{equation}

The total surface of oil is not directly deducible from the mass fraction as some oil droplets may coalesce or evolve through Oswald ripening for example. Thus the initial droplets size can change. The low angle Porod regime however can be used to get an experimental evaluation of the oil droplets surface. The Porod regime is given by:

\begin{equation}
I(q)q^4=2\pi\Delta\rho^2\frac{S}{V}
\end{equation}

 where $\Delta\rho$ is the scattering length density contrast between oil and water, $S$ is the surface of the oil/water interface and $V$ is the scattering volume. From this equation, it is possible to obtain the total surface of the oil droplets.
The surface coverage is then obtained by assessing the excess scattering due to the NPs. To obtain this value, we will compute:
\begin{equation}
Q=\int I(q)q^2dq
\end{equation}
The value of $Q$ computed using the experimental scattering intensity is proportional to the sample composition:
\begin{equation}
Q_{tot}=2\pi^2[\phi_o\phi_w(\rho_o-\rho_w)^2+\phi_o\phi_s(\rho_o-\rho_s)^2+\phi_w\phi_s(\rho_w-\rho_s)^2]
\end{equation}
with $\phi_o$, $\phi_w$ and $\phi_s$ the volume fractions of oil, water and NPs. $\rho_o$, $\rho_w$ and $\rho_s$ are the scattering length densities of oil, water and NPs. If only the extrapolated low angle Porod regime is used to compute $Q$, then we obtain:

\begin{equation}
Q_P=2\pi^2\phi_o\phi_w(\rho_o-\rho_w)^2
\end{equation}

Making the difference between $Q_{tot}$ and $Q_P$, we obtain:
\begin{equation}
Q_{tot}-Q_P=2\pi^2\phi_s[\phi_o(\rho_o-\rho_s)^2+\phi_w(\rho_w-\rho_s)^2]
\end{equation}

knowing the volume fraction of oil from the measured transmission, it is possible to obtain the volume fraction of NPs. As the size of the NPs is precisely known and because no free NPs are present in the sample, it is possible to deduce from the value of the volume fraction the surface coverage of the oil droplets.
\bibliographystyle{unsrt}
\bibliography{Bib_Pickering}

\end{document}